\documentclass[letterpaper, 10 pt, conference]{IEEEtran}

\usepackage{tabularx}

\usepackage{algpseudocode}

\usepackage{algorithm}
\usepackage{algorithmicx}

\usepackage{subfigure}
\usepackage{lipsum} % For algorithm smape after or before:
\usepackage{arydshln}
\usepackage[dvips]{color}
\usepackage{comment}
\usepackage{todonotes}
\usepackage{epsf}
\usepackage{epsfig}
\usepackage{times}
\usepackage{epsfig}
\usepackage{graphicx}
\usepackage{bbold}
\usepackage{mathtools}
\usepackage{mathrsfs}
\usepackage{amssymb}
\usepackage{pdfpages}
\usepackage{epstopdf}
\usepackage{amsmath}
\usepackage{dsfont}
\usepackage{lettrine} % \lettrine[findent=1pt]{{{R}}}{}

\usepackage{amsmath,epsfig,amssymb,algpseudocode,amsthm,cite,url}
\usepackage{caption}

\allowdisplaybreaks
\usepackage{csquotes}
\usepackage{verbatim}
\usepackage[english]{babel}
\usepackage{amsmath,amssymb}

\usepackage{multirow}
\usepackage{rotating}
\usepackage{glossaries}

\usepackage[justification=centering]{caption}
\usepackage{verbatim}

\makeglossaries

\begin{document}
	 \title{Deep Reinforcement Learning for Dynamic Spectrum Sharing of LTE and NR\vspace{-0.5cm}}
	\IEEEoverridecommandlockouts
	\author{\IEEEauthorblockN{Ursula Challita and David Sandberg\\}
			Ericsson Research, Stockholm, Sweden\\
           Email: \{ursula.challita and david.sandberg\}@ericsson.com

%\vspace{-1cm}

	}
	\maketitle
	
\IEEEpeerreviewmaketitle

\newglossaryentry{lte}{name=LTE, description={Long-Term Evolution}}
\newglossaryentry{nr}{name=NR, description={New Radio}}
\newglossaryentry{tti}{name=TTI, description={Transmit Time Interval}}

\vspace{-1cm}	
\begin{abstract}
In this paper, a proactive dynamic spectrum sharing scheme between 4G and 5G systems is proposed. In particular, a controller decides on the resource split between \gls{nr} and \gls{lte} every subframe while accounting for future network states such as high interference subframes and multimedia broadcast single frequency network (MBSFN) subframes. To solve this problem, a deep reinforcement learning (RL) algorithm based on Monte Carlo Tree Search (MCTS) is proposed. The introduced deep RL architecture is trained offline whereby the controller predicts a sequence of future states of the wireless access network by simulating hypothetical bandwidth splits over time starting from the current network state. The action sequence resulting in the best reward is then assigned. This is realized by predicting the quantities most directly relevant to planning, i.e., the reward, the action probabilities, and the value for each network state. Simulation results show that the proposed scheme is able to take actions while accounting for future states instead of being greedy in each subframe. The results also show that the proposed framework improves system-level performance. %Quality of Service (QoS).% in terms of latency and throughput. % \ucc{(i guess we only look at reward. rephrase this...)} while also enhancing overall spectral efficiency.
%In dynamic spectrum sharing, two RATs share the same spectrum and may use all the resources. For LTE-NR DL sharing, LTE scheduler loans resources during a certain time to NR where NR avoids symbols used in LTE for cell specific signals. The controller decides on the resource split between NR and LTE every TTI, and for both UL and DL.
\end{abstract}
\vspace{-0.1cm}
\section{Introduction}
\vspace{-0.1cm}
Dynamic spectrum sharing (DSS) has emerged as an effective solution for a smooth transition from 4G to 5G by introducing 5G systems in existing 4G bands without hard/static refarming spectrum~\cite{EricssonDSS_2}. Using DSS, 4G LTE~\cite{LteOverview} and 5G NR~\cite{NrOverview} can operate in the same frequency band where a controller distributes the available spectrum resources dynamically over time between the two radio access technologies (RATs). For instance, in LTE-NR downlink (DL) sharing, LTE scheduler loans resources during a certain time to NR and NR avoids symbols used in LTE for cell specific signals. Moreover, DSS helps ease the transition from non-standalone 5G networks to standalone 5G. That said, it is important to investigate an effective scheme for the bandwidth (BW) split between LTE and NR to reap the benefits of DSS.

While some literature has recently studied the problem of spectrum sharing between LTE and WiFi (i.e., LTE-unlicensed)~\cite{LTE-U}, NR and WiFi (i.e., NR-unlicensed)~\cite{NR-U}, aerial and ground networks~\cite{UAVcoex}, and radars and communication systems~\cite{radar}, the performance analysis of 4G/5G DSS remains relatively scarce~\cite{EricssonDSS}. For instance, an instant spectrum sharing technique at subframe time scale has been proposed~\cite{EricssonDSS}. The proposed scheme takes into account several information about the cell, such as the amount of data in the buffer, thus splitting the BW between 4G and 5G in every transmission time internal (TTI). Despite the promising results, this work considers a reactive spectrum sharing approach that does not account for the future network states and thus resulting in performance degradation. On the other hand, in a proactive approach, rather than reactively splitting the BW based on incoming demands and serving them when requested, the network takes into account future states for 4G/5G spectrum sharing thus improving the overall system level performance. %guaranteeing \ucc{difficult to "guarantee" QoS, rephrase?} the promised quality of services (QoS) of the users. % and improving spectral efficiency.
%\ucc{mention literature on AI planning, sequential decision making, MCTS, ....}

The main contribution of this paper is to introduce a novel model-based deep reinforcement learning (RL) based algorithm for DSS between LTE and NR. The main scope of the proposed scheme is planning in the time domain whereby the controller distributes the communication resources dynamically over time and frequency between LTE and NR at a subframe level while accounting for future network states over a specific time horizon. To enable an efficient planning, we propose a deep RL technique based on Monte Carlo Tree Search (MCTS)~\cite{MctsSurvey}. When a model of the environment is available, algorithms like AlphaZero~\cite{AlphaZero} have been used with great success. However, in the case of DSS, the LTE and NR schedulers are part of the environment, and these are not easily modelled. Inspired by the MuZero work~\cite{MuZero}, we use a learned model of the environment for planning in the time domain. When applied iteratively, the proposed solution predicts the quantities most directly relevant to planning, i.e., the reward, the action probabilities, and the value for each state. This in turn enables the controller to predict a sequence of future states of the wireless network by simulating hypothetical communication resource assignments over time starting from the current network state and evaluating a reward function for each hypothetical communication resource assignment over the time window. As such, the communication resources in the current subframe are assigned based on the simulated hypothetical BW split action associated with maximized reward over the time window. To our best knowledge, this is the first work that exploits the framework of deep RL for DSS between 4G and 5G systems. Simulation results show that the proposed approach improves quality of service in terms of latency. Results also show that the proposed algorithm results in gain in different scenarios by accounting for several features while planning in the time domain, such as multimedia broadcast single frequency network (MBSFN) subframes and diverse user service requirements.%, and periodic high interference.% during specific subframes.

The rest of this paper is organized as follows. Section~\ref{system_model} presents the system model. Section~\ref{algorithm} describes the proposed deep RL algorithm. In Section~\ref{results}, simulation results are analyzed. Finally, conclusions are drawn in Section~\ref{conclusion}.

\vspace{-0.15cm}
\section{System Model}\label{system_model}
\vspace{-0.1cm}
Consider the downlink of a wireless cellular system composed of a co-located cell operating over NR and LTE serving a set $\mathcal{J}$ of $J$ users. NR and LTE are assumed to operate in the 3.5 GHz frequency band and apply FDD as the duplexing method. We consider a 15 kHz NR numerology and that LTE and NR subframes are aligned in time and frequency. Each RAT, $s$, serves a set of $K_s \subseteq J$ of UEs. The total system bandwidth, $B$, is divided into a set $\mathcal{C}$ of $C$ resource blocks (RBs). Each RAT, $s$, is allocated a set $C_s \subseteq C$ of RBs, and each UE $j \in K_s$ is allocated a set $C_{j,s} \subseteq C_s$ of $C_{j,s}$ RBs by its serving RAT $s$.

For an efficient spectrum sharing model of LTE and RAT, one must design a mechanism for dividing the available bandwidth for data and control transmission for each of the RATs. For the control region, we consider the following:
\begin{itemize}
  \item LTE PDCCH is restricted to symbols \#0 and \#1 (if NR PDCCH is present).
  \item NR has no signals/channels in symbols \#0 and \#1.% (four port LTE CRS is assumed).
  \item NR PDCCH is limited to symbol 2, assuming that the UE only supports type-A scheduling (no mini-slots).
  \item In LTE subframes where no NR PDCCH is transmitted in the overlapped NR slots, LTE PDCCH could span 3 symbols.
\end{itemize}
For data transmission, a controller decides on the resource split, $C_s$, between NR and LTE every subframe. %transmission time interval (TTI). %, for both UL and DL transmissions.
%\begin{figure}[!h]
%\centering
%\includegraphics[width=2.5in]{figures/system_model}
%\caption{Illustration of the system model.}
%\label{fig:system_model}
%\end{figure}

%as depicted in Figure \ref{fig:system_model}.

\vspace{-0.1cm}
\subsection{Channel Model}
We assume the 3GPP Urban Macro propagation model \cite{propagationModel} with Rayleigh fading. The path loss between UE $j$ at location $a$ and its serving BS $s$, $\xi_{j,s,a}$, is given by \textit{Model1} \cite{TR36814}, considering 3.5 GHz frequency band:% with the constant term recalculated for 3.5 GHz (instead of 2.0 GHz):
\begin{equation}\label{pathloss_equation}
\xi_{j,s,a}= 20.4 + 37.6\times \textrm{log}_{10}(d),
\end{equation}
where d is the distance between the UE and the BS in meters. The signal-to-noise ratio (SNR), $\Gamma_{j,s,c,a}$ of the UE-BS link between UE $j$ at location $a$ served by RAT $s$ over RB $c$ will be:
\begin{equation}\label{SNR_equation}
\Gamma_{j,s,c,a}= \frac{P_{j,s,c,a} h_{j,s,c,a}}{B_c N_0},
\end{equation}
where $P_{j,s,c,a} = \overline{P}_{j,s,a}/C{j,s}$ is the transmit power of BS/RAT $s$ to UE $j$ at location $a$ over RB $c$ and $\overline{P}_{j,s,a}$ is the total transmit power of BS/RAT $s$ to UE $j$ location $a$. Here, the total transmit power of RAT $s$ is assumed to be distributed uniformly among all of its associated RBs. $h_{j,s,c,a} = g_{j,s,c,a}10^{-\xi_{j,s,a}/10}$ is the channel gain between UE $j$ and BS/RAT $s$ on RB $c$ at location $a$ where $g_{j,s,c,a}$ is the Rayleigh fading complex channel coefficient. $N_0$ is the noise power spectral density and $B_c$ is the bandwidth of an RB $c$. Therefore, the achievable data rate of UE $j$ at location $a$ associated with RAT $s$ can be defined as:
\begin{equation}\label{rate}
R_{j,s,a}=\sum_{c=1}^{C_{j,s}}B_c \textrm{log}_2(1+\Gamma_{j,s,c,a}),
\end{equation}

\subsection{Traffic Model}
We assume a periodic traffic arrival rate per UE $j$ with a fixed periodicity $\lambda_j$ and a fixed packet size $\beta_j$. Time domain scheduling is typically governed by a scheduling weight whereby a high weight corresponds to a high priority for scheduling that particular UE. We adopt a similar mechanism for measuring the quality of bandwidth splits between LTE and NR where a UE not fulfilling its QoS is associated with a high weight. % As such, the delay weight, $w_{p,j}(t)$, of user $j$ during subframe $p$ can be expressed as the weight of the most delayed packet per user $j$ during that subframe.
The weight for user $j$ in subframe $p$ can be calculated as:

\vspace{-0.15cm}
\begin{equation}\label{delay_weight}
w_{p,j}(t)=  \begin{cases}
      \alpha_jt & t< \delta_j, \\
      \alpha_jt + \eta_j & t\geq \delta_j.
   \end{cases}
\vspace{-0.1cm}
\end{equation}

where $t$ is the time the oldest packet has been waiting in the buffer, $\delta_j$ and $\eta_j$ correspond to the step delay and step weight of the delay weight function of user $j$, respectively and $\alpha_j$ is a small positive factor that makes the weight non-zero when there is data in the buffer. Note that a UE with zero weight will not be scheduled. Here, the step delay corresponds to the maximum tolerable delay in order to maintain QoS. If a packet remains in the buffer for a time period larger than $\delta_j$, the weight for user $j$ increases by $\eta_j$.

%\subsection{Broadcast signalling}
%In DSS, it is important to account for synchronization and broadcast signals that are sent in LTE and NR. Those signals are called PBCH/SS in LTE and SSB in NR and are necessary for the UEs to find the cell and get access to the network. In the spectrum sharing, the LTE PBCH/SS and NR SSB signals are allocated in different subframes and the subframes that send the NR SSBs are configured as MBSFN subframes. As such, the NR SSB signal will not be interfered by the LTE signals such as LTE CRS. Some of these signals will be allocated to a set of PRBs (like LTE PBCH) while others (like NR SSBs) will require configuration of MBSFN subframes~\cite{LteOverview}.

Given this system model, next, we develop an effective spectrum sharing scheme that can allocate the appropriate bandwidth to each RAT, at a subframe time scale, while accounting for future network states.

\section{Deep Reinforcement Learning for Dynamic Spectrum Sharing}\label{algorithm}
In this section, we propose a proactive approach for DSS enabling LTE and NR to operate on the same BW simultaneously. In this regard, we propose a deep RL framework that enables the controller to learn the BW split between LTE and NR during subframe $t$ while accounting for future network states over a time window $T$. To realize that, first, we propose the adopted RL algorithm for training the controller to learn the optimal policy for BW split. Then, we introduce the RL architecture and components for the DSS problem.

\subsection{Deep RL Algorithm}
To enable a proactive BW split between LTE and NR, we adopt in this paper the MuZero algorithm~\cite{MuZero}.
One of the main challenges of the proposed solution technique is that it requires a model for the individual schedulers for LTE and NR, which is hard to devise. Instead, we propose in this paper to learn the scheduling dynamics via a model-based reinforcement learning algorithm that aims to address this issue by simultaneously learning a model of the environment's dynamics and planning with respect to the learned model~\cite{MuZero}. This approach is more data efficient compared to model-free methods where current state-of-the-art algorithms may require millions of samples before any near-optimal policy is learned.

During the training phase of the proposed algorithm, the prediction comprises performing a MCTS over the action space and over the time window $T$ to find the sequence of actions that maximizes the reward function. MCTS iteratively explores the action space, gradually biasing the exploration towards regions of states and actions where an optimal policy might exist. To enable our model to learn the best explored sequence of actions for each network state, we define three neural networks - the representation function ($h$), dynamics function ($g$), and prediction function ($f$). The motivation for incorporating each of these neural networks in the proposed algorithm is described as follows:
%\ucc{Just to remember: $p$: current (real) subframe number . $i$ level in the search tree (i=0 means root node). }
\begin{itemize}
  \item \emph{A representation function ($h$)} encodes the observation in subframe $p$ into an initial hidden state ($s_0$).
  \item \emph{A dynamics function ($g$)} computes a new hidden state ($s_{i+1}$) and reward ($r_{i+1}$) given the current state ($s_i$) and an action ($a_{i+1}$).% The dynamics function mirrors the structure of a Markov Decision Process (MDP) model that computes the expected reward and state transition for a given state and action.
  \item \emph{A prediction function ($f$)} outputs a policy ($p_i$) and a value ($v_i$) from a hidden state ($s_i$).
\end{itemize}
During the training phase, the model predicts the quantities most directly relevant to planning, i.e., the reward, the action probabilities and the value for each state. The proposed training algorithm is summarized in Algorithm~\ref{training_algorithm} and the main steps are given as follows:
\begin{itemize}
  \item \emph{Step 1}: The model receives the observation of the network state as an input and transforms it into a hidden state ($s_0$).
  \item \emph{Step 2}: The prediction function ($f$), is then used to predict the value $v_i$ and policy vector $p_i$ for the current hidden state $s_i$.
  \item \emph{Step 3}: The hidden state is then updated iteratively to a next hidden state $s_{i+1}$ by a recurrent process consisting of $N_{\textrm{unroll}}$ steps, using the dynamics function ($g$), with an input representing the previous hidden state $s_i$ and a hypothetical next action $a_{i+1}$, i.e., a communications resource assignment selected from the action space comprising allowable bandwidth splits between LTE and NR.
  %\item \emph{Step 4}: The model is subsequently unrolled recurrently for $K$ steps. At each step $k$, the dynamics function $g$ receives as input the hidden state $s_{k+1}$ from the previous step and the hypothetical action $a_{k}$.
  \item \emph{Step 4}: Having defined a policy target, reward and value, the representation function ($h$), dynamics function ($g$), and prediction function ($f$) are trained jointly, end-to-end by backpropagation-through-time (BPTT).
\end{itemize}
%\ucc{shall we add anything about data generation?}
Meanwhile, the testing algorithm refers to the actual execution of the algorithm after which the weights of ($h$), ($g$), and ($f$) have been optimized and is implemented for execution during run time. Given that DSS is performed on a 1 ms basis, it is too demanding to run MCTS online. As such, we use the representation ($h$) and prediction ($f$) functions only during test time. The main steps performed by the controller at test time are summarized in Algorithm~\ref{execution_phase}.

\begin{algorithm}[t!] \scriptsize
\caption{Training phase}\label{training_algorithm}
\begin{algorithmic}[t!]
\vspace{0.2cm}
\State \textbf{Input:} Representation ($h$), dynamics ($g$) and prediction ($f$) functions.
\For {$i$ = 0 ... $N_{\textrm{iter}}-1$}
\vspace{0.1cm}
\State \textbf{\small{Data Generation}}
\vspace{0.1cm}
%\State \textbf{Input:} Representation ($h$), dynamics ($g$) and prediction ($f$) functions.
%\State \textbf{Output:} Replay buffer $\mathscr{R}$.
\State \textbf{Step 1:} Sample $N_{\textrm{episode}}$ environments ($\mathscr{E}$) with random parameters.% and initialization.
\For {($env$ $\in$ $\mathscr{E}$)}
%\For {($t$ $\in$ execution time)}
\For {$p$ = 0 ... $N_{\textrm{timestep}}-1$}
\State \textbf{Step 2:} Encode the observation $o_p$ into an initial hidden state, $s_0$.
\State \textbf{Step 3:} Run $N_{\textrm{mcts}}$ MCTS simulations from this state using ($g$) and ($f$).
\State \textbf{Step 4:} Sample an action to take in the environment.% with probabilities proportional to the number of times each hypothetical action $a_x$ has been taken during simulation.
\State \textbf{Step 5:} Store the policy ($\pi$) and reward ($u$) in the replay buffer ($\mathscr{R}$).%, where $a_i$ is the action, $\pi_i$ is the search policy and $r_i$ is the received reward.
%\State \textbf{Step 5:} A tuple ($a_i$, $p_i$, $r_i$) is stored in a replay buffer ($\mathscr{R}$), where $a_i$ is the action, $\pi_i$ is the search policy and $r_i$ is the received reward.
\EndFor
\EndFor
\vspace{0.1cm}
\State \textbf{\small{Neural Network Training}}
\vspace{0.1cm}
\For {$p$ = 0 ... $N_{\textrm{step}}-1$}
%\State \textbf{Step 6:} Sample $M$ sequences from the replay buffer $\mathscr{R}$, where $M$ is the batch size.
\State \textbf{Step 6:} Sample a batch of sequences from the replay buffer $\mathscr{R}$.%, where $M$ is the batch size.
\State \textbf{Step 7:} Compute the total discounted reward ($z$) over each sequence.
\State \textbf{Step 8:} Take a training step using BPTT to make $p$ $\approx$ $\pi$, $v$ $\approx$ $z$, $r$ $\approx$ $u$.
%\State \textbf{Step 7:} Take a training step using BPTT with the dynamics function unrolled for $N_{\textrm{unroll}}$ time steps to make $p$ $\approx$ $\pi$, $v$ $\approx$ $v$, $r$ $\approx$ $r$.
\EndFor
\EndFor
\end{algorithmic}
\end{algorithm}

\begin{algorithm}[t!] \scriptsize
\caption{Execution phase}\label{execution_phase}
\begin{algorithmic}[t!]
\vspace{0.2cm}
\State \textbf{Input:} Representation ($h$) and prediction ($f$) functions.
\For {$n$ = 0 ... $N_{\textrm{timestep}}-1$}
%\For {($t$ $\epsilon$ execution time)}
\State \textbf{Step 1:} Encode the observation into an initial hidden state, $s_0$.
%\State \textbf{Step 1:} At time $t$, the observation is encoded into an initial hidden state, $s_0$, via the representation function ($h$).
\State \textbf{Step 2:} Calculate the action probabilities using ($f$) and select the best action.%Select the action with the highest action probability to take as $argmax pPredict the The prediction function ($f$) predicts the best action, based on the maximum policy, for time $t$.
%\State \textbf{Step 2:} The prediction function ($f$) predicts the best action, based on the maximum policy, for time $t$.
\State \textbf{Step 3:} Find the BW split to use and send that to the schedulers.
%\State \textbf{Step 3:} The controller is informed which action to take regarding bandwidth split between the two RATs for time $t$ and the decision is sent to each of the schedulers.
\EndFor
\end{algorithmic}
\end{algorithm}
%The main MCTS loop iterates over $num_simulations$, where one simulation is a pass through the MCTS tree until a leaf node (i.e. unexplored node) is reached and subsequent backpropagation

\subsection{Deep RL Components}
In this subsection, we define the RL framework components, namely the observations, actions, and rewards.

\begin{itemize}
  \item \textbf{Action space:} BW split between LTE and NR for DL transmission for subframe $p$, denoted as $a_p=\{a_1, a_2, ... a_N\}$ where $N$ is the size of the action space. Here, an action corresponds to a horizontal line splitting the BW on one side to LTE and the other side to NR. The possible BW splits are chosen by grouping a set of multiple RBs thereby resulting in a quantized action set. This would in turn reduce the action space size and is valid due to the fact that the gain between bandwidth splits from consecutive RBs is negligible. %The total number of possible actions is $n^T$, where $T$ is the future time window and $n$ is the total number of possible bandwidth splits at a particular TTI.
  \item \textbf{Observation:} the observation for subframe $p$, denoted as $o_p$, is divided into two parts, where the first part, ($o_{p,1}$), consists of components with size $(J\textrm{x}1)$ whereas the second part, ($o_{p,2}$), consists of components with size $(J\textrm{x}T)$, where $T$ is the time window consisting of a set of future subframes. The different observations components are summarized as follows: %. Below is a summary of the different observation elements:
  \begin{itemize}
    \item NR support: a vector with $J$x1 elements that indicates if a user $j$ is NR user or not.%, $\in o_{p,1}$
	\item Buffer state: a vector with $J$x1 elements containing the number of bits in the buffer of user $j$.%, $\in o_{p,1}$
   % \item Number of arrived packets: a vector with $J$ elements containing the number of received packets in the buffer of user $j$ during a particular episode.
    \item MBSFN subframe: a matrix with $J$x$T$ elements that indicates for each subframe $p$, $p\in T$, if a UE is configured with MBSFN or not. By configuring LTE UEs with MBSFN subframes, some broadcast signalling can be avoided at the cost of decreased scheduling flexibility.%, $\in o_{p,2}$
    \item Predicted number of bits per PRB and TTI for each UE $j$: a matrix with $J$x$T$ elements, where each element contains the estimated number of the average bits that can be transmitted for user $j$ in subframe $p$, taking into account the estimated channel quality of user $j$ during subframe $p$, $p\in T$.%, $\in o_{p,2}$% as well as possibly reserved resources in that PRB. Hence, a PRB that is used for e.g., LTE PBCH would have zero estimated bits for all users. Similarly, an MBSFN subframe would have zero estimated bits for all LTE UEs.
    \item Predicted packet arrivals: a matrix with $J$x$T$ elements indicating the number of bits that will arrive in the buffer for each user $j$ over a set of future subframes $T$.%, $\in o_{p,2}$
  \end{itemize}
  \item \textbf{Reward function:} the reward function is modelled as a summation of the exponential of the most delayed packet per user and can be expressed as:
      \begin{equation}\label{reward}
        r_p=e^{-\sum_{j=1}^{J} w_{p,j}},
      \end{equation}
      where $w_{p,j}$ is the delay weight function of user $j$ in subframe $p$, as described in (\ref{delay_weight}). The intuition behind this reward function is that high total weight is penalized with a low reward in subframe $p$. Meanwhile, if the controller manages to keep the user buffers empty, the reward per subframe will be one. If a highly prioritized UE is queued for several subframes, its weight will increase and thus the reward will approach zero.

\end{itemize}
\begin{figure}[!t]
\centering
\includegraphics[width=2.5in]{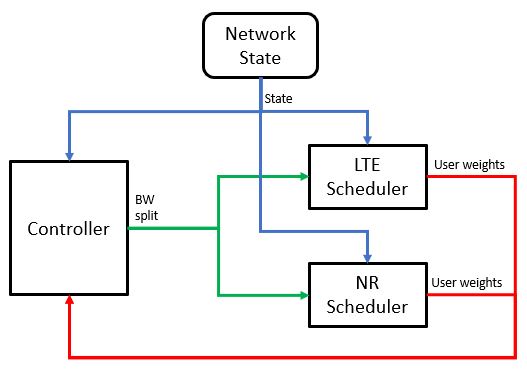}
\caption{A schematic illustration of the proposed setup summarizing the connection between the controller, network state, and LTE and NR schedulers.}\label{fig_setup}
\vspace{-0.2cm}
\label{fig:controller}
\end{figure}

Figure \ref{fig_setup} summarizes the relationship between the network state, controller, and LTE and NR schedulers. At each subframe, the LTE scheduler, NR scheduler, and controller receive the network state information. This information is then used by the controller to generate observations and thus take an action for the BW split between LTE and NR. This action is then conveyed to the LTE and NR schedulers. Given the network state information and the corresponding BW split, each of the schedulers allocates their respective users to the corresponding BW portion for the current subframe. Finally, the weights for the users are fed to the controller and used as an input for the calculation of the reward. Next, we provide simulation results and analysis for the proposed RL framework.

\begin{table}[t!]\footnotesize
\setlength{\belowcaptionskip}{0pt}
\setlength{\abovedisplayskip}{3pt}
\captionsetup{belowskip=0pt}
\newcommand{\tabincell}[2]{\begin{tabular}{@{}#1@{}}#1.6\end{tabular}}
 \setlength{\abovecaptionskip}{2pt}
 \renewcommand{\captionlabelfont}{\small}
 \captionsetup{justification=centering}
\caption{Simulation parameters for the radio environment.}\label{simulation_parameters_radio}
\centering
\tabcolsep=0.03cm %to reduce table width
\begin{tabular}[t]{|c|c|}
\hline
\textbf{Parameter} & \textbf{Value} \\
\hline
Frequency & 3.5 GHz\\
\hline
Bandwidth & 25 PRBs (5 MHz) \\
\hline
Traffic Model & Periodic\\
\hline
UE speed & 3 m/s\\
\hline
Transmit power & 0.8W/PRB \\
\hline
Noise power ($N_0$) & 112.5 dBm/PRB \\
\hline
Antenna config & 1 Tx, 2 Rx \\
\hline
\end{tabular}
\vspace{-0.24cm}
\end{table}

\section{Simulation Results and Analysis}\label{results}
In this section, we provide simulation results and analysis for the performance of the proposed algorithm under four different scenarios where planning in the time domain for dynamic spectrum sharing is relevant. Tables \ref{simulation_parameters_radio} and \ref{simulation_parameters_NN} provide a summary of the main simulation parameters.

\begin{figure}[!t]
\centering
\includegraphics[width=2.7in]{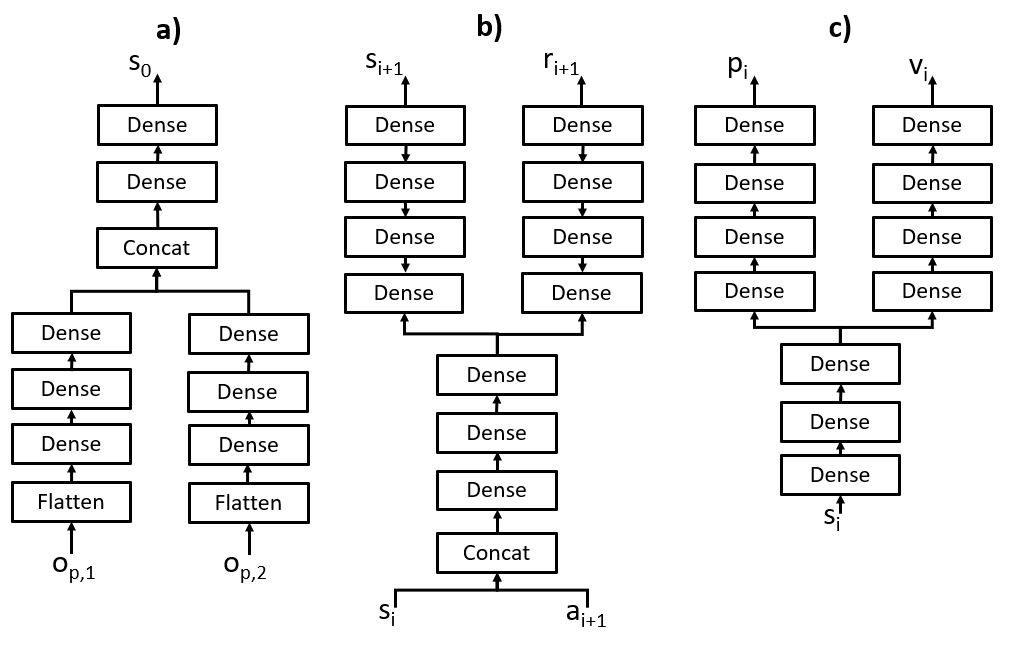}
\caption{Neural network architectures for a) the representation function, b) the dynamics function and c) the prediction function.}
\vspace{-0.3cm}
\label{fig:neural_nets}
\end{figure}

The structure of the representation, dynamics, and prediction neural networks is depicted in Figure \ref{fig:neural_nets}. All dense layers except for the output layer use 64 activations with ReLU activation. The representation outputs ($s$) use 10 activations with $tanh$ activation. The reward ($r$) and value ($v$) outputs are scalar with linear activation, and the policy ($p$) has the same number of activations as the number of actions with $softmax$ activation.

\begin{table}[t!]\footnotesize
\setlength{\belowcaptionskip}{0pt}
\setlength{\abovedisplayskip}{3pt}
\captionsetup{belowskip=0pt}
\newcommand{\tabincell}[2]{\begin{tabular}{@{}#1@{}}#1.6\end{tabular}}
 \setlength{\abovecaptionskip}{2pt}
 \renewcommand{\captionlabelfont}{\small}
\captionsetup{justification=centering}
\caption{Simulation parameters for the RL framework.}\label{simulation_parameters_NN}
\centering
\tabcolsep=0.03cm %to reduce table width
\begin{tabular}[t]{|c|c|}
\hline
\textbf{Parameter} & \textbf{Value} \\
\hline
Number of MCTS simulations ($N_{\textrm{mcts}}$)  & 64 \\
\hline
Episode length ($N_{\textrm{timestep}}$)  & 16 subframes \\
\hline
Discount factor ($\gamma$) & 0.99 \\
\hline
Window size (T) & 10 subframes \\
\hline
Batch size & 32 examples \\
\hline
Number of unroll steps ($N_{\textrm{unroll}}$)& 3 \\
\hline
Number of TD steps ($N_{\textrm{td}}$)  & 16 \\
\hline
Optimizer & Adam \\ %\ref{kingma2014adam}
\hline
Learning rate & $1*10^{-4}$ \\
\hline
Number of episodes per iteration ($N_{\textrm{episode}}$) & 100 \\
\hline
Representation size & 10 \\
\hline
\end{tabular}
\vspace{-0.24cm}
\end{table}

%The simulation setup consists of two identical schedulers for LTE and NR users respectively. A prior step (called Arbitratior) decides the available frequency resources for each scheduler.

\begin{algorithm}[t!] \scriptsize
\caption{Baseline Algorithm}\label{baseline_algorithm}
\begin{algorithmic}[t!]
\vspace{0.1cm}
\State \textbf{Input:} Observation, $o_t$.
\State \textbf{Output:} Action, $a_t$.
\vspace{0.1cm}

\State \textbf{Step 1:} Calculate the weight of each user according to Eq \ref{delay_weight}.
\State \textbf{Step 2:} Sort the users in order of decreasing weight.
\State \textbf{Step 3:} Schedule users from the list until spectrum is full.
\State \textbf{Step 4:} Check how many PRBs are needed for NR and LTE users.
\State \textbf{Step 5:} Select the action, $a_t$, that splits the BW proportionally between the RATs.
%\State \textbf{Step 5:} Select the action, $a_t$, that splits the BW proportionally between LTE and NR users.
\end{algorithmic}
\end{algorithm}

Next, we provide a detailed description for the simulation results and analysis of each of the four studied scenarios. Note that in all of the scenarios, the episode length is 16 and thus the evaluation score for a perfectly solved scenario is also 16. Moreover, we assume that LTE users (if any) are scheduled on the lower part of the spectrum band and NR users (if any) are scheduled on the high part of the band. As for the baseline, we split the available spectrum proportionally to the number of required RBs between LTE and NR users, as summarized in Algorithm \ref{baseline_algorithm}. We also compare the performance of the proposed algorithm to equal BW split and alternating BW between LTE and NR.
The user weight is calculated using Eq. \ref{delay_weight}, with $\eta_j=5$ and $\alpha_j=10^{-5}$ for all users. The step delay, $\delta_j$, is set appropriately for the different users in the different scenarios as specified below.

%To simplify the problem we assume that the arrival of future traffic can be perfectly predicted and that .... For some types of traffic (like periodic ...) this is not too unrealistic. Also future SINRs are assumed to be perfectly predicted. (\ucc{do we need to mention these here? maybe better to clarify these points when we are describing the observation vector?})

\subsection{Scenario 1: MBSFN subframes}
LTE requires CRSes to enable demodulation of data. Therefore, if only NR UEs are scheduled, the CRSes are not needed and are hence an overhead. If there is a lot of NR traffic to be scheduled, LTE can be configured with so called MBSFN subframes. In these subframes, no CRSes are transmitted and it is therefore not possible to schedule LTE users but this can result in improved efficiency for NR users. This scenario aims to investigate if the controller can learn to account for MBSFN subframes during planning thus enabling time critical LTE traffic to be served before MBSFN subframes.

%            weightCalculator = 'Delay(stepDelay=3, stepWeight=5, slope=0.00001)'
%            if userId==0:  # NR 0:
%                trafficModel = 'Periodic(periodicity=4, packetSize=45000, startOffset=0.5)'
%            elif userId==1:  # LTE 1:
%                trafficModel = 'Periodic(periodicity=4, packetSize=15000, startOffset=0.5)'

\subsubsection{Scenario description}
We consider two users, one NR user and one LTE user, both having a traffic arrival periodicity of 4 ms and a step delay $\delta_0$=3ms. The packet size is 45000 bits and 15000 bits for the NR and LTE users, respectively. The system is configured with a repeating MBSFN pattern with a periodicity of 4 subframes, where the first two subframes are non-MBSFN (i.e., both LTE and NR UEs can be scheduled) and the last two subframes in the pattern are MBSFN subframes (i.e., only NR UEs can be scheduled).

\subsubsection{Optimal bandwidth split}
To solve this scenario optimally, both packets must be served within 3 ms. As such, the LTE user should be served in the non-MBSFN subframes to make resources available for the NR user later in the cycle. Therefore, the optimal strategy is to start scheduling LTE such that its buffer is emptied before the MBSFN subframes.

\subsubsection{Results and analysis}
\begin{figure}[!h]
\vspace{-0.2cm}
\centering
\includegraphics[width=2.5in]{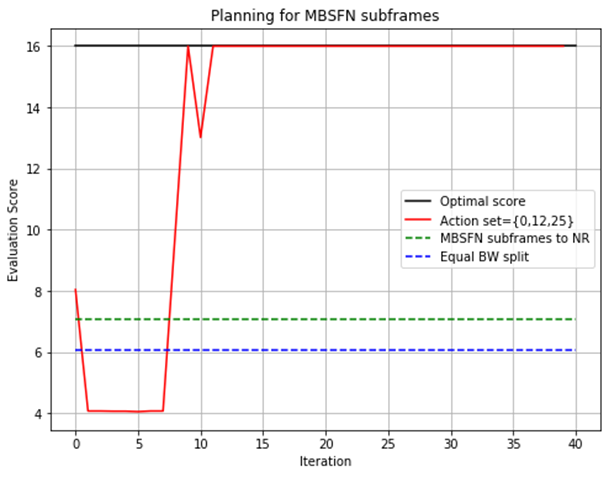}
\caption{Evaluation score as a function of number of iterations for scenario 1 with MBSFN subframes.}
\vspace{-0.4cm}
\label{fig:MBSFN}
\end{figure}

From Figure \ref{fig:MBSFN}, we can see that the proposed algorithm converges to the optimal strategy in 12 iterations. Also, note that the performance of the proposed scheme exceeds that of equal bandwidth split between LTE and NR and the case where MBSFN subframes are allocated to the NR user and non-MBSFN subframes are allocated to the LTE user. With this MBSFN configuration the amount of overhead due to e.g. LTE CRSes can be minimized which results in improved efficiency on network level. The controller can learn to account for the MBSFN subframes by scheduling in such a way that maximizes the quality of service despite the reduced scheduling flexibility due to the MBSFN subframes.

%        elif userMode==4: #periodical interference
%            if userId==0:  # NR 0:
%                trafficModel = 'Periodic(periodicity=2, packetSize=14000, startOffset=0.5)'
%                weightCalculator = 'Delay(stepDelay=2, stepWeight=2, slope=0.00001)'
%            elif userId==1:  # LTE 1:
%                trafficModel = 'Periodic(periodicity=2, packetSize=10000, startOffset=0.5)'
%                weightCalculator = 'Delay(stepDelay=2, stepWeight=2, slope=0.00001)'

\subsection{Scenario 2: Periodic high interference}
In this scenario, we investigate the controller's ability to learn to account for future high interference on one of the users during planning. Periodic high interference can, for instance, occur in case a user is at the cell edge and is interfered by another base station or in case of unsynchronized time division duplexing scenarios.

\subsubsection{Scenario description}
We consider two users, one NR user and one LTE user, both having a traffic arrival periodicity of 2 ms.
We assume a larger packet size for NR user compared to that of LTE so that we can observe the gain of NR benefiting from the 2 extra symbols of LTE PDCCH if it is allocated the full bandwidth. Users have a small weight value when the delay is less than 2 ms but then it increases abruptly to 2 after 2 ms (i.e., $\delta=2ms$). Moreover, a periodic high interference is observed on LTE user every 3 subframes. Here, the periodic interference term is added artificially for analysis purposes.

\subsubsection{Optimal bandwidth split}
The optimal strategy for this scenario is to allocate the full bandwidth to NR during subframes with high interference on the LTE user.

\subsubsection{Results and analysis}
\begin{figure}[!h]
\vspace{-0.2cm}
\centering
\includegraphics[width=2.7in]{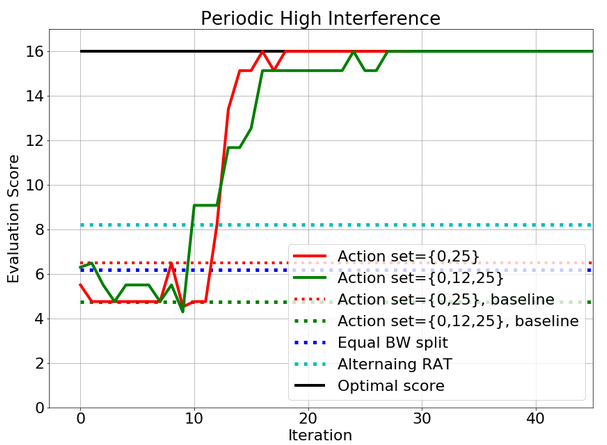}
\caption{Evaluation score as a function of number of iterations for scenario 2 with periodic high interference.}
\vspace{-0.4cm}
\label{fig:periodic_interference}
\end{figure}

From Figure \ref{fig:periodic_interference}, we can see that the proposed algorithm converges to the optimal strategy in 18 and 28 iterations for the case of 2 and 3 action space, respectively. The proposed approach outperforms the baseline algorithm, equal bandwidth split, and alternating bandwidth split where the controller learns to allocate the full bandwidth to NR during subframes with high interference for the LTE user as opposed to taking actions based on buffer status only. This allows the controller to split the bandwidth between LTE and NR such that the impact of the interference level from neighboring cells is reduced thus resulting in an improved system level performance.

%The red and the green curve corresponds to the performance of the controller with two available actions (0 or 25 PRBs to NR) and three available actions (0, 12 or 25 PRBs to NR) respectively. The red dashed and the green dashed curve corresponds to the performance of the baseline algorithm described by Algorithm \ref{baseline_algorithm} with different action sets. The blue dashed curve corresponds to an equal bandwidth split in each subframe. The cyan curve corresponds to a policy where the full bandwidth is allocated to NR in even subframes and to LTE in odd subframes.

\subsection{Scenario 3: Mixed services}
In this scenario, we investigate the controller's ability to handle users with different delay requirements.
\subsubsection{Scenario description}
We consider two users, one high priority NR user ($\delta_0=5ms$) with 90000 bits, and one low priority LTE user ($\delta_1=10ms$) with 90000 bits. Data arrives in subframe 1 for both users.

%        elif userMode==11: #mixed services (modified for paper)
%            if userId==0:  # NR 0:  Initial medium prio NR user
%                trafficModel = 'Periodic(periodicity=15, packetSize=90000, startOffset=0.5)'
%                weightCalculator = 'Delay(stepDelay=5, stepWeight=10, slope=0.00001)'
%            elif userId==1:  # LTE 0: Initial low prio LTE user
%                trafficModel = 'Periodic(periodicity=15, packetSize=90000, startOffset=0.5)'
%                weightCalculator = 'Delay(stepDelay=10, stepWeight=10, slope=0.00001)'
%            else:
%                raise ValueError('Too many users for this scenario')

\subsubsection{Optimal bandwidth split}
The optimal strategy for this scenario is to postpone the scheduling of the low priority LTE user in order to allow the high priority NR user to be scheduled. When the buffer of the high priority users is emptied, the controller can schedule the LTE user.

\subsubsection{Results and analysis}
\begin{figure}[!h]
\vspace{-0.2cm}
\centering
\includegraphics[width=2.5in]{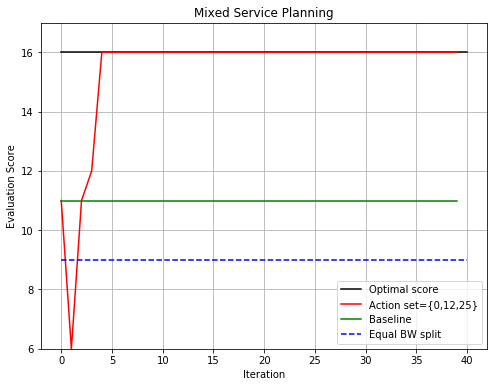}
\caption{Evaluation score as a function of number of iterations for scenario 3 with mixed services.}
\vspace{-0.4cm}
\label{fig:mixed_services}
\end{figure}

From Figure \ref{fig:mixed_services}, we can see that the proposed approach converges to the optimal policy within 5 iterations. The controller learns to prioritize the NR user with a tight delay requirement over the LTE user thus outperforming the baseline algorithm as well as the equal BW split.

\subsection{Scenario 4: Time multiplexing}
In this scenario, we investigate the controller's ability to learn to do time multiplexing (as opposed to frequency multiplexing) between LTE and NR. Time multiplexing can result in two extra symbols for NR when no LTE is scheduled due to the fact that no LTE PDCCH needs to be transmitted. This in turn results in an increased efficiency when the RATs are scheduled in a time multiplexed fashion.

\subsubsection{Scenario description}
We consider two users, one NR user and one LTE user, both having a traffic arrival periodicity of 2 ms.
The packet size for the NR user is larger (14000 bits) compared to that of the LTE user (10000 bits). Users have a small weight when delay is less than 2 ms but then increases abruptly to 5 after 2 msec (i.e. $\delta=2ms$).

\subsubsection{Optimal bandwidth split}
The optimal strategy for this scenario is to perform time multiplexing whereby the full bandwidth is allocated to a particular RAT every other subframe. As such, NR could benefit from the 2 extra symbols of LTE PDCCH when it is given the full bandwidth. This results in a larger transport block size and thus the large NR packet size can be served within one subframe.

\subsubsection{Results and analysis}
\begin{figure}[!h]
\vspace{-0.2cm}
\centering
\includegraphics[width=2.7in]{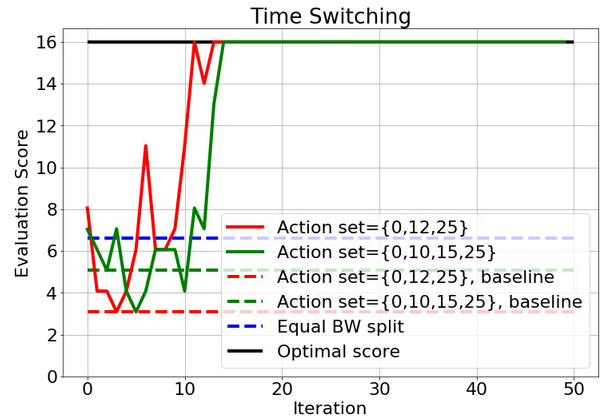}
\caption{Evaluation score as a function of number of iterations for scenario 4 considering a time multiplexing scenario.}
\vspace{-0.4cm}
\label{fig:time_switching}
\end{figure}

From Figure \ref{fig:time_switching}, we can see that the proposed approach converges to the optimal action strategy within 14 and 15 iterations for the case of 3 and 4 actions, respectively. The proposed approach outperforms the baseline algorithm, equal BW split, and alternating BW split where the network learns to perform time multiplexing between LTE and NR resulting in an increased spectrum efficiency. For the studied scenario, when NR is scheduled alone, i.e. without overhead from LTE PDCCH, the maximum transport block size is 14112 bits. On the other hand, when LTE is scheduled with NR, there is an extra overhead for LTE PDCCH and thus the maximum transport block size decreases to 12576 bits. Consequently, the NR packet can be scheduled in one subframe given that NR is scheduled alone during that subframe.

%The red and the green curve corresponds to the performance of the controller with three available actions (0,12 25 PRBs to NR) and four available actions (0, 10, 15 or 25 PRBs to NR) respectively. The red dashed and the green dashed curve corresponds to the performance of the baseline algorithm described by Algorithm \ref{baseline_algorithm} with the previous action spaces. The blue dashed curve corresponds to an equal bandwidth split in each subframe.

\section{Conclusion}\label{conclusion}
In this paper, we have proposed a novel AI planning framework for dynamic spectrum sharing of LTE and NR. Results have shown that the controller can split the bandwidth between LTE and NR in an intelligent way while accounting for future network states, such as MBSFN subframes and high interference level, thus resulting in an improved system level performance. This gain comes from the fact that the proposed algorithm uses knowledge (or beliefs) about future network states to make decisions that perform well on a longer timescale rather than being greedy in the current subframe. As part of future work, we aim to further investigate if the suggested algorithm can learn to account for uncertainties in the observations.% as well as scenarios with MBB traffic that can show gain with planning.}
%Another advantage is that since the weights can be treated as a cost of delaying traffic rather than the urgency of scheduling a user, configuring the weight functions for different types of traffic becomes much more straight-forward.

%with MBB traffic it makes sense to use some other quality metric to optimize, for example proportional fair

%study how the algorithm behaves in more realistic episode lengths that are close to unlimited

%\newpage
%To select the next best move, it makes sense to ‘play out’ likely future scenarios from the current position, evaluate their value using a neural network and choose the action that maximises the future expected value.

%The UCB score is a measure that balances the estimated value of the action Q(s,a)with a exploration bonus based on the prior probability of selecting the action P(s,a) and the number of times the action has already been selected N(s,a). Early on in the simulation, the exploration bonus dominates, but as the total number of simulations increases, the value term becomes more important.

%these are predicted rewards, rather than actual rewards from the environment.

%We use Monte Carlo estimation of the target value (i.e. the discounted sum of all future rewards to the end of the episode)

\def\baselinestretch{0.87}
\bibliographystyle{IEEEtran}
\bibliography{references}
\vspace{-0.5cm}
\end{document}